# Quantitative measurement of magnetic dichroic signals at sub-nanometer to atomic scale resolution conditions

Sharath Kumar Manjeshwar Sathyanath[1], Anna L. Ravensburg[2], Vassilios Kapaklis[2], Jan Rusz[2], Björgvin Hjörvarsson[2], Klaus Leifer[1] *

1 Department of Materials Science and Engineering, Uppsala University, Box 35, 75237 Uppsala, Sweden
2 Department of Physics and Astronomy, Uppsala University, Box 516, 75237 Uppsala, Sweden
* Corresponding author

**Abstract**

When aiming for atomic resolution electron magnetic circular dichroism (EMCD) in STEM mode, the high convergence angle of the electron probe can lead to unforeseen artefacts and strong reductions of the signal to noise ratio (SNR) in the measurement of magnetic magnitudes. In this work, the EMCD signal is measured in STEM mode at semi-convergence angles ranging from $\alpha = 2 - 10$ mrad using iron samples with high structural perfection. We observe that the relative EMCD signal remains very similar for all convergence angles, which is in good agreement with simulations. The improvement in the signal-to-noise ratio as compared to earlier works is due to better crystalline quality of the Fe sample, high control of the sample orientation, better control over sample thickness, and factors such as higher beam current and the use of a fast and sensitive hybrid-pixel electron detector. One of the key factors to remove experimental ambivalences is the use of an acquisition geometry where the two conjugate EMCD spectra are acquired simultaneously. Furthermore the high crystalline quality Fe sample enables the acquisition of statistical entities of spectra at the same sample orientation, which in turn results in high SNR EMCD signals. We obtain high SNR EMCD signals for all semi convergence angles and could measure the orbital to spin magnetic moment ratio $m_l/m_s$ with high precision. We observed that at high convergence angles, the $m_l/m_s$ ratio gets sensitive to minute changes in sample orientation.

## Introduction

Magnetic properties of materials play a pivotal role in a wide range of applications, including devices for data storage and sensors. Understanding and controlling the magnetic properties at the nanoscale and atomic scale is crucial for developing next-generation materials and devices. Electron magnetic circular dichroism (EMCD) is a technique that can probe at these length scales and measure the magnetic moments at high spatial resolution [1], [2], [3], [4]. Classical EMCD, although powerful, is inherently associated with a low signal-to-noise ratio, which is a significant drawback. Furthermore, in the original proposal of the EMCD technique, a close to parallel beam geometry is used yielding only low spatial resolution in the analysis. The two main advantages of the EMCD technique, as compared to XMCD is that it operates close to atomic scale as well as it allows for site selection of the signal by steering electron channelling. It has been shown that using a convergent beam in scanning mode, the signal-to-noise ratio (SNR) can be maintained at nanometer scale as well as atomic resolution can be potentially achieved [5], [6], [7]. For instance, Thersleff et al. demonstrated the detection of an EMCD signal in convergent mode at a semi-convergence angle of 8 mrad, corresponding to a probe size of 2 Å, thereby overcoming the spatial limitations of the classical EMCD technique [5]. Wang et al. reported atomic-plane-resolved EMCD measurements by utilizing chromatic aberration correction in parallel beam illumination using close to parallel beam illumination [6]. Ali et al. successfully improved the technique through the use of double apertures as well as they measured the EMCD signal on a zone axis at a semi-convergence angle of 10 mrad by using ventilator entrance apertures [7], [8], [9]. The latter could yield an EMCD signal at convergence angles corresponding to atomic resolution conditions, though the zone axis orientation leads to a complex situation of the inelastic scattering. There have been few experimental studies on improving SNR and spatial resolution in STEM mode [5] at one or two convergence angles. When increasing the convergence angle to reduce the electron probe size, not only the number of beams corresponding to the formation of the EMCD signal, increases drastically, but also direct interference is present where the convergent beam disks overlap. Thus, when approaching atomic resolution in EMCD measurements, it is important to understand how the measured EMCD signal, its SNR and the $m_l/m_s$ ratio obtained from the sum-rule formula evolve with convergence angle. Here, we report a systematic experimental study of the evolution of the EMCD signal as a function of semi-convergence angle on an Fe samples that show an extremely small rocking curve angle as well as by acquiring the data on a state of the art hybrid electron camera. The experimental conditions in this work enable the systematic study of EMCD signal, its SNR and the $m_l/m_s$ ratio obtained from the applications of sum-rules and further a quantitative understanding and interpretation of the EMCD signal at high convergence angles.

Experimental details

For this EMCD study, it is important to have an Fe crystal with highest quality and with a thickness of 20 nm at which a very good EMCD signal is expected. Using magnetron sputtering, 20 nm thick Fe films were deposited on a $MgAl_2O_4$ substrate. Further experimental details of the deposition process are reported in [10], [11], [12]. As a result, the rocking curve angle of those samples is as small as ±0.014 °. Since the sample orientation is very stable over a large area, we can compare and accumulate data coming from different scan points. This study can thus rely on the acquisition of several 10.000 spectra used to improve the statistical quality as well as to reduce impact of beam damage on the analysis.

Plan view samples were prepared by polishing a 3 mm disc of the sample to a thickness of about 100 μm. Subsequently, a dimple was created on the substrate side of the polished disc using a GATAN dimple grinder, with a central area of approximately 20 μm to 25 μm. This was followed by grazing incidence argon ion polishing (GATAN PIPS II), starting at 5 kV and lowering the voltage to 4, 3 kV and 2 kV until perforations were formed. The final polishing was done at 1.2 kV. Then, the sample was cleaned with acetone, followed by isopropanol. Subsequent plasma cleaning was carried out before being loading the specimen into the TEM.

The sample was analysed in STEM mode using a Titan/Themis probe-corrected TEM at an acceleration voltage of 200 kV, fitted with a CEOS CEFID energy filter equipped with a direct electron detector, ELA. Spectra of the $L_3$ and $L_2$ edges of Fe, corresponding to the positions C+ and C- in the diffraction plane, see Figure 1, were acquired at the 2-beam condition. The direct electron detector has an excellent SNR, where in the area outside where there is no signal on the camera, there are only occasional single counts. For the case of this analysis this means that the SNR of one spectrum with acquisition time t is the same or very similar to the SNR of a sum spectrum taken over a large number of spectra with the same total acquisition time t. This advantage is key when approaching atomic resolution in STEM mode where the dwell time at one pixel is limited by beam damage.

The EMCD signal appears in the half plane above and below the 0 - g-line in the inelastic diffraction pattern at energy losses corresponding here to the Fe-$L_3$ and Fe-$L_2$ edge, see Figure 1. Two spectra from the areas denominated C+ and C- are acquired and their difference spectrum is calculated. When accurate EMCD measurements are to be carried out, it is important to acquire both spectra simultaneously. The reason for this is that during the first scan at the C+ position, the electron beam damage might buckle the sample and contamination can appear. In addition, we believe that an accurate positioning of the electron beam with respect to the atomic planes is needed for obtaining well defined EMCD spectra, and due to sample drift, the exact register between the atomic positions at the first and second scan maybe lost. In many cases, cross-correlation will not establish the correct register again or leave doubts on the accuracy of this process. To eliminate such artefacts, a slit at the entrance aperture position of the spectrometer enables the simultaneous acquisition of C+ and C- spectra at each scan point. The slit aperture can be positioned on the wished $q_x$ direction (Figure 1). Since the spectrometer is set in spectroscopy mode, called qE mode, on its camera, energy loss is projected along the x-direction, i.e. on one point of the spectral trace all intensities along the $q_x$ direction and inside the slit width in x-direction are summed. Electrons inelastically scattered along the $q_y$ direction are collected along the y-direction. Therefore, by acquiring such camera image, we keep the freedom to position the digital aperture for the extraction of the C+ and C- signal along the $q_y$ direction in the evaluation process.

The EMCD experiments were carried out with the sample oriented such that the (110) reflections were excited. For each α (α= 2 mrad – 10 mrad), a 100 x 100 pixels map was acquired, where the qE map was recorded at each pixel, thus resulting in a 4D EMCD data set. For these convergence angles, the probe size ranges from 1.7 Å - 8 Å The typical dose per scan point (nm$^2$) varies from 2.2 $e^7$ electrons/pixel to 1.09 $e^9$ electrons/pixel.

In this work, we adapted a methodology that enables high control on sample orientation, thickness and iron oxidation. Therefore, at every convergence angle, a high energy-loss dataset including the Fe-L edge and the O-K edge with collection window centred on the 0-beam, a low-loss dataset to measure thickness, and a 4D dataset consisting of a diffraction scan to evaluate the sample orientation. Each acquisition contains 2000-10,000 spectra. This facilitates the analysis of individual or sum spectra and the use of statistical methods for data analysis.

The data was evaluated using HyperSpy [13], an open-source Python package for hyperspectral data treatment. The C+ and C- traces were extracted from the original 4D data and converted into two 3D data sets. These 3D datasets (C+ and C-) comprise thus of two spatial dimensions (2D) and one signal dimension (1D). The background subtraction was performed by selecting a pre-edge window of approximately 50 eV, spanning from 650 eV to 700 eV. The background-subtracted data is post-edge normalized by selecting a window of approximately 30 eV, spanning from 730 eV to 760 eV. The EMCD signal was extracted by taking the difference of C+ and C- spectra. The data were oversampled in the energy dispersion direction (0.01 eV to 0.1 eV) to calibrate the C+ and C- spectra. The $m_l/m_s$ ratio was calculated using EMCD sum rules [14].

Simulations of inelastic scattering were performed by the MATS.v2 method [15] using a convergence parameter set to $10^{-5}$, using combined multislice and Bloch-waves method [16]. Convergence angles were set to values from 1 mrad to 25 mrad, with a step of 1 mrad. Beam was centered on an atomic plane. Electron beam was tilted from the (001) zone axis to the (1,-1,8) direction, i.e. approximately 10 degrees, to reach a systematic row orientation with G=(110). Beam was further tilted along the systematic row orientation to reach various conditions spanning between the 3-beam orientation (zero tilt) to 2-beam orientation (tilt of G/2). Scattering cross-section was calculated on a grid spanning from -25 mrad to +25 mrad with a step of 1 mrad along the both $\theta_x, \theta_y$, scattering directions. A supercell of bcc iron containing 132 atoms was constructed with a supercell c-axis along the (1,-1,8) direction and a-axis along the (1,1,0) direction [17]. Sample thickness was set to 20 nm.

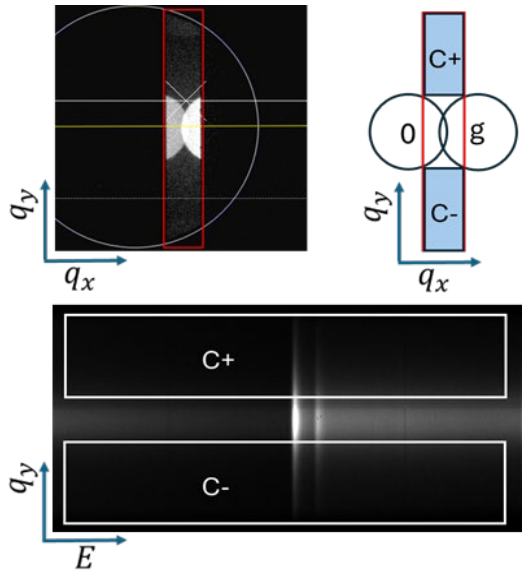

Figure 1: Representative image of the position of the slit aperture for $q_x$, $q_y$ selection (left upper), the schematic of the inelastic diffraction pattern together with the positions C+ and C- at which the EMCD signal appears (right upper) and the obtained spectral trace on the camera from which C+ and C- Fe edges are extracted (lower figure).

Results

Figure 2a shows the annular dark field (ADF) image of the scanned region of the sample where the substrate has been removed. Most of the region of interest shows a bright contrast indicating an orientation close to 2-beam-condition (2BC). The relative thickness of the scanned area of about 0.2 t/λ was calculated from the low loss spectrum in Figure 2b. In the convergent beam electron diffraction (CBED)pattern in Figure 2c, the g-beam, i.e. the (110)-reflection has the strongest intensity. From the 4D diffraction scans, the relative intensity $\eta_g$ ($\eta_g = I_g / I_0$) was extracted for all x-y pixels in the scan to measure possible changes of the sample orientation as further discussed below.

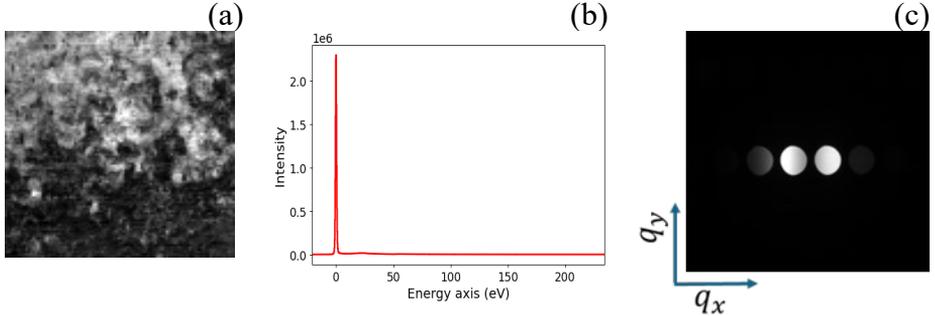

Figure 2: a) ADF image of the scanned area as well as corresponding zero-loss peak (ZLP) (b) and the CBED pattern showing the diffraction condition with the 0 and (110) beams excited (c).

The EMCD results for a 5 mrad semi-convergence angle are presented in Figure 3. The raw spectra collected from the plus (C+) and minus (C-) positions are shown in Figure 3 (a). In Figure 3 (b), the background-subtracted and post-edge-normalized spectra are overlaid, revealing a clear difference at the $L_3$ and $L_2$ edges. The EMCD signal, which is the difference between C+ and C- spectra, is displayed in the enlarged plot along with the fit for the $L_3$ and $L_2$ difference signals. The integral over the fitted $L_3$ and $L_2$ difference signals are denominated $dL_3$ and $dL_2$ respectively. In the following, $dL_3$ and $dL_2$ will also be called the $L_3$ and $L_2$ EMCD signal. The area of $dL_3$ and $dL_2$ are taken to calculate the $m_l/m_s$ ratio in (c). The very good SNR in those difference spectra enables us to quantitatively evaluate changes of the EMCD signals as well as of the $m_l/m_s$ ratio.

First, we analyse how the EMCD signal strength changes with the convergence angle. In Figure 4 (a) the EMCD difference signal for all semi-convergence angles is shown. One observes that the SNR in these difference spectra remains excellent for all convergence angles. This means that using a highly convergent electron beam, the EMCD signal contains quantitative information and, seen the similarity of the difference spectra, should be exploitable to obtain and analyse the $m_l/m_s$ ratio.

The relative $L_3$ and $L_2$ EMCD signals are defined as $rL_{3/2} = dL_{3/2}/L_{3/2}$. In Figure 4(b) one observes that an excellent strength of the relative EMCD signal is maintained until a semi-convergence angle of 10 mrad. The $rL_3$ signal has a value of 5.6 % for a convergence angle of 2 mrad, then reaches values of 8 % at 3-4 mrad and finally decreases to 4.5-5 % at semi-convergence angles of 9-10 mrad. The $rL_2$ value decreases in a similar way for convergence angles between 3-6 mrad, and remains at a stable level of $rL_3 = 4.1 - 4.7$ % at $\alpha = 7 - 10\ mrad$. This rather high stability of the relative EMCD signal for high convergence angles is surprising, because the EMCD signal arises from the interference of electron waves related to the zero beam and the **g**-beam that have a well-defined vector product and phase relation. When the electron beam is converged as compared to a parallel beam illumination, a much more complex interference situation is present and intuitively, one might expect a clearer decrease of the EMCD signal, especially for the high semi-convergence angles, where the convergent beam discs clearly overlap. Thus, this is good news for future quantitative EMCD studies at atomic resolution using a focused electron beam. In the following, we analyse the impact of high convergence angle on quantitative aspects of the EMCD analysis down to atomic resolution conditions.

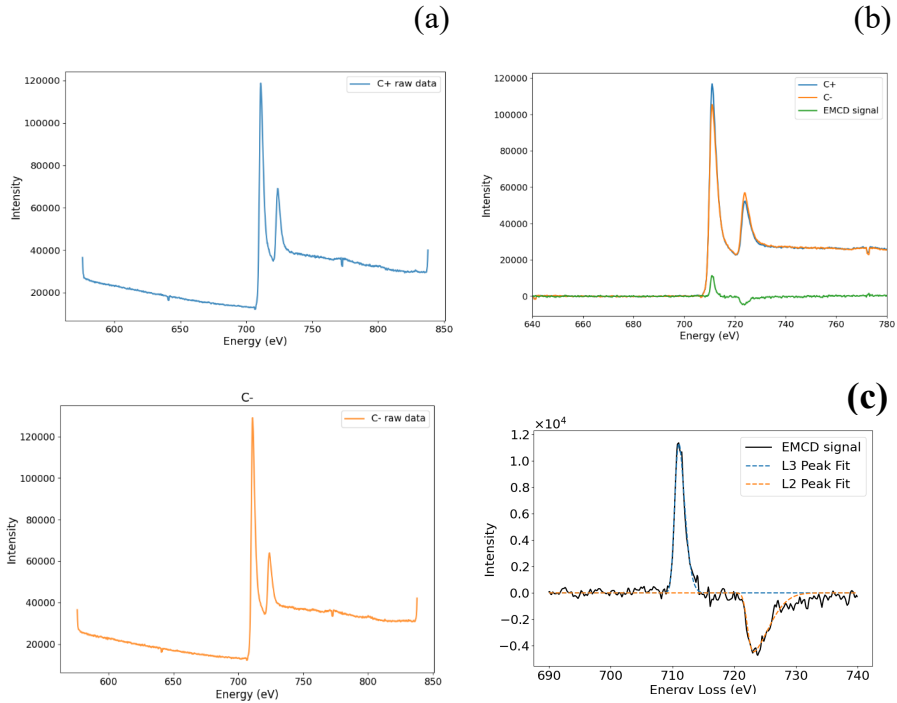

Figure 3: Plots from α=5 mrad dataset. (a) C+ and C- raw spectra. The minima at about 640 eV and 770eV are related to stitching pixels where the different camera segments meet. (b) shows the background subtracted and post-edge normalized C+ and C- spectra along with the EMCD difference signal(c) showing EMCD signal with fit used to quantify $m_l/m_s$ ratio.

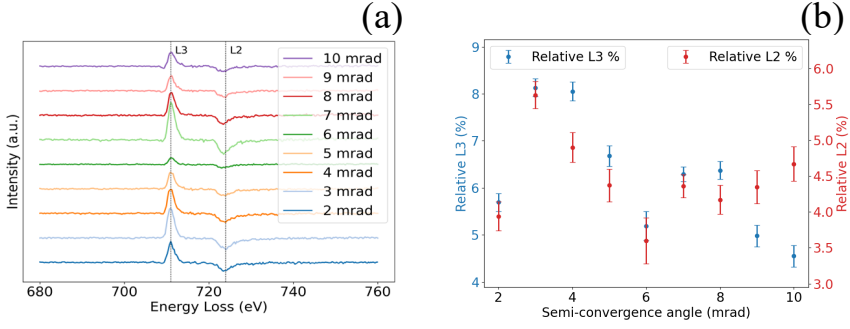

Fig. 4: a) EMCD difference spectra for all α = 2 mrad -10 mrad, b) evolution of relative $L_3$ signal and the relative $L_2$ signal as a function of semi-convergence angle.

In the next step, we simulated the relative EMCD signal $rL_3$ for different convergence angles. The simulations in Figure 5 for the relative EMCD signal were carried out at 5 sample orientations ranging from the (110) 2BC to the (110) 3BC. In all orientations, a clear decrease of $rL_3$ is observed with increasing semi-convergence angle with the exception of the (110) 3BC orientation, where $rL_3$ increases for smaller α.

In the (110) 2BC condition, $rL_3$ decreases by about a factor of 2 for $\alpha$ varying between 3 – 10 mrad in good correspondence to the experimental observation. When the orientation is closer to the 3BC, for example at 0.25x2BC and at the 3BC itself, $rL_3$ varies only very little as a function of $\alpha$.

When the orientation of the sample would change, from this Figure 5, we expect a stronger change in $rL_3$ for small $\alpha$ than at larger semi-convergence angles. The smaller value of the $rL_3$ for $\alpha = 2\ mrad$ as compared to the one for $\alpha = 3\ mrad$ could potentially be explained by a slight overall orientation difference in the scanned area. Though, we took care to keep the scanned area the same for all convergence angles. From a careful observation of the ADF images corresponding to the scans at 2 mrad and 3 mrad, we can observe a minor sample drift. Though, the overall $I_g/I_0$ intensity, taken from the zero loss diffraction pattern, is high and similar for the scans taken at both semi-convergence angles. This indicates an overall very stable 2BC in both scans. Therefore, we cannot explain here the deviation of the $rL_3$ at $\alpha = 2\ mrad$ signal from the expected, higher value here.

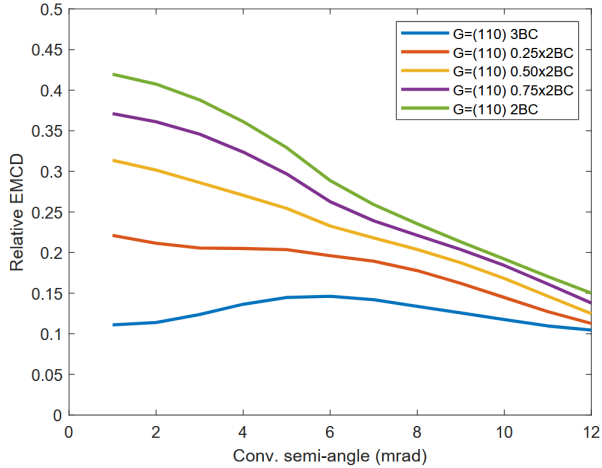

Figure 5: Simulation of the relative EMCD signal $rL_3$ as a function of the semi-convergence angle for several sample orientations indicated in the figure legend.

To assess the orientation of the scanned area in more detail, the average diffraction pattern from the 100×70 pixel region, along with its profile for 4 mrad semi-convergence angle is illustrated in Figure 6. The profile indicates that the average $\eta_g$ is considerably high. The intensity levels closely match those documented in earlier studies by Ali et al.[18], where the orientation was very close to the 2BC.

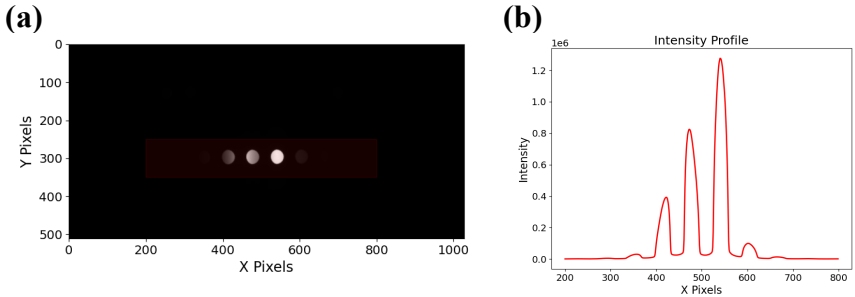

Figure 6: Diffraction pattern with Fe (110) 2BC excited for 4 mrad convergence angle along with the intensity profile through 0 beam and (110) beam.

To further assess the impact of sample orientation on the magnitudes extracted from EMCD measurements, we have evaluated the orientation maps, represented by $\eta_g$. In Figure 7(a) we show the map corresponding to the data for the 4 mrad semi-convergence angle. These maps contain the upper 70 % of the region of interest in Figure 2a, i.e. the part of the ADF image that has a bright contrast indicating an orientation close to 2BC. The $I_0$ and $I_g$ intensities are extracted from discs that include the entire CBED disc for small α. When the CBED discs start to overlap, $I_0$ and $I_g$ intensities are extracted from the part where the discs do not overlap. The map shows that the sample maintains a stable orientation within the 100×70-pixel region. The $\eta_g$ values are significantly higher for low α. This is expected since for small α, the incoming beams are closer to the 2BC. The mean values for $\eta_g$ in Figure 7b thus decrease by a factor of 2.5 over the range of semi-convergence angles chosen in this work. In the following, we analyse on how the sample orientation impacts the EMCD signal. The accurate knowledge of the orientation dependence of $rL_3$ is one of the most important parameters in the EMCD experiment.

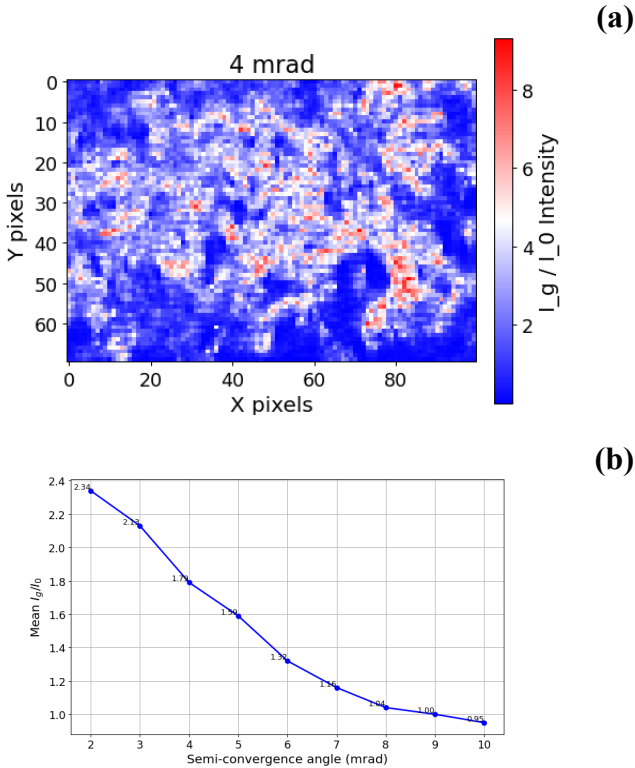

Fig. 7: a) $\eta_g$ orientation maps and b) mean value of $\eta_g$ as a function of α.

It has been shown for small convergence angles in [18] that the $m_l/m_s$ ratio remains stable for sample tilts in the range of 2 mrad. One of the exciting and unique measurements with EMCD consists in the characterisation of magnetic interfaces at atomic resolution. In this case, in particular, the relative EMCD signals, $rL_{3/2}$ gives important information since changes of this value would indicate local changes in the atomic magnetisation. In the following, we analyse the $rL_3$ changes as a function of orientation for the different α.

On the Fe samples, in this work, the rocking curve angle is 0.014°= 0.2 mrad. Thus, in the scanned area in this work, the orientation variation between the Fe domains is of this order or smaller. In principle, to obtain the change of $rL_3$ with sample orientation, one could calculate the standard deviation of the $rL_3$ obtained from all 7000 spectra in one 4D data set. But, due to the small SNR of the individual spectra, the noise would strongly modify the changes of $rL_3$ and dominate the observed standard deviation of $rL_3$ values. This means that the change of $rL_3$ with change of orientation would most likely be lost in the SNR of the individual spectra, especially for the change of orientation between different Fe domains that is about 10 times smaller than the one of the Fe film in the work [18].

Therefore, we choose a different way to determine the $rL_3$ change as a function of sample orientation. We first distribute the $\eta_g$ values in the maps in Figure 7a into channels in histograms, where every channel corresponds to an interval of $\eta_g$ values, see Figure 8a. We calculate the histogram for every second semi-convergence angle. Then, we map the corresponding EMCD spectra into those channels and calculate the average value of $rL_3$ for each channel. Subsequently, we calculate the weighted standard deviation of $rL_3$ from each histogram corresponding to one angle α. The values of the standard deviation in Figure 8b range from 0.4 % at 2 mrad to 3 % at 8 mrad. This means that very small changes in crystal orientation of the order of 0.2 mrad clearly change the $rL_3$ value. This curve equally indicates that $rL_3$ is more sensitive to orientation changes for higher convergence angles.

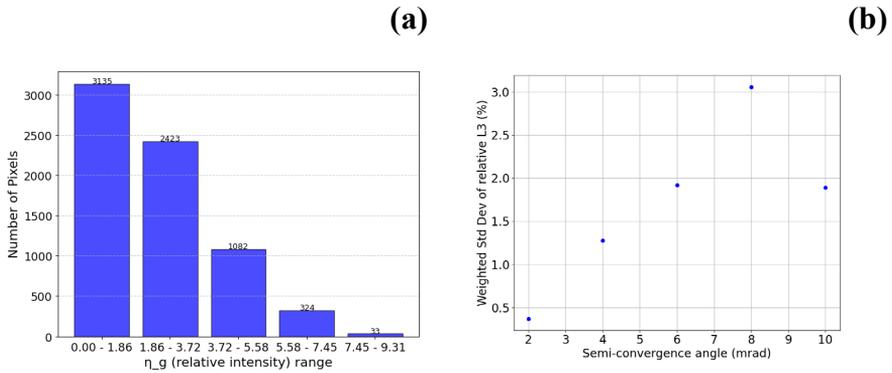

Figure 8: (a) Distribution of $\eta_g$ for 4 mrad semi-convergence angle and (b) weighted standard deviation of $rL_3$ from each histogram corresponding to one angle α.

The most common magnetic magnitude extracted from circular dichroic data is the $m_l/m_s$ value which was calculated based on the sum rules [14] using the equation given below and the values $dL_3$ and $dL_2$. Table 1 presents the $m_l/m_s$ values for various semi-convergence angles, which is in good agreement with previously reported EMCD experiments [19], [20].

$$\frac{m_l}{m_s} = \frac{2}{3} \frac{\int_{L3} \Delta I\,(E)\,dE + \int_{L2} \Delta I\,(E)\,dE}{\int_{L3} \Delta I\,(E)\,dE - 2\int_{L2} \Delta I\,(E)\,dE}$$

| Semi-convergence angle (mrad) | $m_l / m_s$ |
|---|---|
| 10 | 0.109 ± 0.007 |
| 9 | 0.067 ± 0.007 |
| 8 | 0.095 ± 0.007 |
| 7 | 0.091 ± 0.007 |
| 6 | 0.041 ± 0.009 |
| 5 | 0.060 ± 0.007 |
| 4 | 0.045 ± 0.007 |
| 3 | 0.060 ± 0.007 |
| 2 | 0.049 ± 0.007 |

Table 1: $m_l/m_s$ values for the different convergence angles.

Discussion

The quantification of the orbital-to-spin magnetic moment ratio ($m_l/m_s$) from our datasets, acquired under a 2BC geometry at different semi-convergence angles yielded values close to the Fe magnetic moments reported by XMCD and gyro-magnetic measurements.[21] Some of the minor differences in $m_l/m_s$ appear for $\alpha = 2 - 6\ mrad$, but they remain within the the statistical error. A comprehensive treatment of sources of noise in such spectra is given in Hu et al [22].

For the $\alpha > 6\ mrad$, the average $m_l/m_s$ ratio is about 0.09, i.e. clearly above the expected value of $m_l/m_s$ = 0.04-0.06. The determination of the $m_l/m_s$ ratio being till to date the key result coming from EMCD analysis of a magnetic sample, we analyse the reason for this discrepancy. From theory, we expect that the $m_l/m_s$ ratio does not change when the beam convergence is varied. Are the $m_l/m_s$ ratios possibly higher because in the complex interference situation arising from scattering of the highly convergent electron beam the assumptions for the sumrules should be modified?

As mentioned above, the relative EMCD signal is most sensitive to the sample orientation for the data acquired at higher α. Having the spectra from the individual channels in histograms in Figure 8a available, we can now determine the $m_l/m_s$ ratio for those spectra that are close to or on the 2BC. In order to obtain a small statistical error of $m_l/m_s$, from the 5 channels in a histogram, we do select the third one, where the signal is averaged over more than 1000 spectra and the $\eta_g$ value is still high. The $m_l/m_s$ ratios reported in table 2 clearly decrease as compared to the $m_l/m_s$ ratios averaged over the entire data set and reach the expected value. From this result, we conclude, that sum-rules, at least for an Fe crystal are still applicable, even for convergence angles as high as 10 mrad, provided, the sample is very well oriented into the 2BC.

| Semi-convergence angle (mrad) | $m_l / m_s$ |
|---|---|
| 8 | 0.06 |
| 10 | 0.05 |

Table 2: $m_l /m_s$ values for the different convergence angles averaged over spectra that were taken from the third channel in the histograms such as shown in Figure 8a.

Conclusion

In this study, we systematically studied how the semi-convergence angle α affects the EMCD signal, aiming to understand how quantitative the different magnitudes extracted from EMCD spectra are and if sum-rules can be used at high convergence angles corresponding to atomic sized electron beams. When increasing the semi-convergence angle, i.e. when changing the probe size, the experimental relative EMCD signal changes in a similar way as theory predicts. When $\alpha$ is in the range where the CBED discs do not overlap, the $m_l/m_s$ ratio is in the range of the values observed in XMCD and gyromagnetic ratio measurements. For higher $\alpha$, we observe both, an increased $m_l/m_s$ ratio as well as an increased change of the EMCD signal with very small orientation changes of the sample in the range of a few 0.2 mrad. Having available the diffraction data for each scan point, we could extract the EMCD spectra at high $\alpha$ value from places where the sample is close to the 2BC. In fact, we could observe that these spectra contain a $m_l/m_s$ ratio that is in the range of the expected values. Thus, by carefully setting the orientation of the magnetic sample, quantitative EMCD work can be carried out at atomic resolution conditions.

Acknowledgements

The computations were enabled by resources provided by the National Academic Infrastructure for Supercomputing in Sweden (NAISS), partially funded by the Swedish Research Council through grant agreement no. 2022-06725. We thank the ARTEMI Swedish national infrastructure for transmission electron microscopy for the support of this work as well as the Swedish national infrastructure Myfab. We thank for the generous support of the Swedish Science Council and the Olle-Engqvist Foundation leading to the acquisition of a new energy filter, whose unique elements made this work possible.

We sincerely thank CEOS for their invaluable support in handling the technical aspects of the spectrometer, as well as for their assistance in installing the slit aperture, which was essential for the EMCD experiments.